\documentclass[prl,
showpacs,
twocolumn,
floatfix
]{revtex4}

\usepackage{graphicx}
\usepackage{bm}
\newcommand{\bq}{\begin{equation}}
\newcommand{\eq}{\end{equation}}

\newcommand{\ev}{\end{verse}}

\newcommand{\bv}{\begin{verse}}

\begin{document}
\title{Non-exponential relaxation and hierarchically constrained dynamics in
a protein}

\author{Elihu Abrahams}
\affiliation{Center for Materials Theory, Serin Physics
Laboratory, Rutgers University, Piscataway, NJ 08854-8019}
\date{\today}
\begin{abstract}

A scaling analysis within a model of hierarchically constrained dynamics is shown to
reproduce the main features of non-exponential relaxation observed in kinetic studies
of carbonmonoxymyoglobin.

\end{abstract}

\pacs{87.15.Da, 64.70.Pf, 78.30.Jw, 82.20.Rp}

\maketitle

There have been various works in which similarities between the dynamics of proteins and the structure and dynamics of glasses and spin glasses have been discussed
\cite{gold, dls, icko, shibata}.
Although the energy landscapes in the two systems may be quite different,
non-exponential relaxation is observed in both. In glasses, relaxation often follows
the stretched exponential form characteristic of the ``Kohlrausch law:"
\bq
\Phi(t) = \Phi_0 \exp[-(t/\tau)^\beta], \;\; 0<\beta<1, 
\eq
where $\tau$ is a temperature-dependent characteristic time which becomes
unmeasurably long as the glass transition temperature is approached. It is
often experimentally observed to follow, over 10 orders of magnitude, a
Vogel-Fulcher law \cite{vf}, $\tau \propto\exp [A/(T-T_0)]$ 

Now, any reasonable relaxation function $\Phi(t)$ can be fit by assuming some
distribution $w(\tau)$ of relaxation times among additive contributions to the
relaxing quantity, thus
\bq
\Phi(t) = \int_0^\infty d\tau w(\tau) \exp(-t/\tau).
\eq
This extends the idea of conventional Debye relaxation with a single
relaxation time to a situation in which there is a distribution of degrees of
freedom each contributing independently to $\Phi(t)$ with its own relaxation
time - thus {\it parallel} relaxation.

A different point of view was proposed by Palmer, Stein, Abrahams, and
Anderson \cite{g4}. They pointed out that the conventional parallel picture, while simple,  is 
often microscopically arbitrary and that a more physical view is that the path to
equilibrium is governed by many sequential correlated steps - thus a {\it
series} interpretation in which there are strong correlations between
different degrees of freedom. These authors proposed \cite {g4} a
microscopically motivated model of hierarchically constrained dynamics (HCD) which
leads to the Kohlrausch law (and a maximal relaxation time of the Vogel-Fulcher form). That result was
cited by Shibata, Kurata, and Kushida \cite{shibata} to argue that HCD holds in their
observation of stretched exponential relaxation in an experiment on conformational
dynamics in Zn-substituted myoglobin. Of course it is possible, and sometimes likely, that both parallel and sequential processes exist in the same system.

In a discussion of anomalous
relaxation in proteins, Iben {\it et al} \cite{icko} presented data on
carbonmonoxymyoglobin (MbCO). This is myoglobin in which CO is bound to the central
iron atom of the heme group. In this system,  parallel relaxation processes also occur; they dominate the dynamics of the rebinding of the ligand after photodissociation \cite{rha}. Here we focus on the pressure
release relaxation experiments of Iben {\it et al}. \cite{icko} Infrared absorption spectra of the stretch bands of the CO were taken under various conditions.  After pressure release,
the center frequency of the $A_0$ band initially shifts rapidly upward by 0.4
cm$^{-1}$ and then relaxes slowly toward its low-pressure equilibrium value, 1.2
cm$^{-1}$ higher. This behavior at various temperatures is shown in Fig.\ 3 of Ref.
1 and it is replotted here (without the error bars) as  Fig.\ 1. It seen that the
relaxation is close to power law over more than three decades of time, thus much
slower than exponential, or even stretched exponential. 

\begin{figure}
\includegraphics[width=8cm]{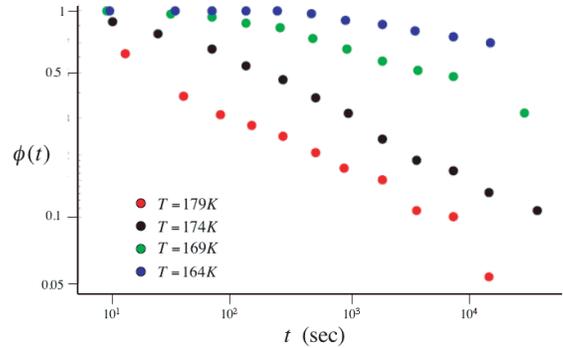}
\caption{Experimental relaxation in MbCO \cite{icko}}
\label{fig1}
\end{figure}

It is of interest to ask whether there is a  model of hierarchically
constrained dynamics as proposed in Ref.\ \onlinecite{g4} which can account for the main
features of Fig.\ 1. These are: a) At each temperature there is a
region of power-law relaxation which crosses over at shorter times to
something much slower. b) This crossover time increases as
temperature decreases. c) The power law is the same at each temperature, but
there is an increasing offset as the temperature is lowered.

In what follows, it will be shown that one of the models mentioned in Ref.\ \onlinecite{g4}
in fact gives behavior identical to what was observed in MbCO. For
HCD, one recognizes that equilibrium
distributions in configuration space are not relevant since the free energy
barriers which determine relaxation are continuously changing in time as
different degrees of freedom relax at different rates. Furthermore, in
a strongly correlated system one expects that with any choice of
coordinates, interactions will remain in the form of constraints and that
these will be of importance over a range of time scales. The nature
of constraints is that some degrees of freedom cannot relax until their
motion is made possible by the relaxation of other degrees of freedom. These
restrictions occur over a wide hierarchy of coordinates, from fast ones to
slow ones. A complete discussion of this HCD approach is given in Ref.\ \onlinecite{g4}.

To implement this picture, Palmer {\it et al}. \cite{g4} set up a hierarchy of
levels $n=0,1,2,3, \ldots$. The degrees of freedom in level $n$ are
represented by $N_n$  Ising pseudospins (two-level centers) each of which has two
possible states. This was adapted from earlier work of Stein \cite{dls}. Constraints enter via the requirement that each ``spin" in
level
$n+1$ is free to change its state only if a condition on some number $\mu_n$
of spins in level $n$ is satisfied.  Now, the
$\mu_n$ spins have $2^{\mu_n}$ states. Let the required condition be that just one of
these possible states is realized. If the average relaxation time  in level
$n$ is $\tau_n$, then on average, it will take a time $2^{\mu_n}\tau_n$ for a
spin in level $n+1$ to change its state. Therefore
\bq
\tau_n = \tau_0 \prod_0^{n-1}\exp (\mu_i \ln 2) = \tau_0\exp U_n,
\eq
where
\bq
U_n = \sum_0^{n-1}\mu_i \ln 2.
\eq
The relaxation function is given as a sum of the correlation functions of all
the degrees of freedom $S_i$:
\bq
\Phi(t) = (1/N)\sum_{i=1}^N \langle S_i(0)S_i(t)\rangle. 
\eq
In a correlated system, the dynamics of the $S_i$ are not independent,
so each of the correlation functions in Eq.\ (5) depends on the
behavior of the other $S$'s. As described above, the HCD scheme of Palmer {\it et
al}.
\cite{g4} incorporates correlations. The
$S_i$ are arranged in a hierarchy of levels
$n$ with each level having its characteristic relaxation time $\tau_n$ given
by Eq.\ (3). So Eq.\ (5) may be rewritten as a sum over the different levels
from 0 to
$\infty$:
\bq
\Phi(t) = \sum_{n=0}^\infty w_n \exp(-t/\tau_n),
\eq
where $w_n = N_n/N$ is the fraction of the total number of degrees of freedom
which are in level $n$. \cite{finite}

For a given model, $w_n$ and $\mu_n$ must be specified. As remarked
in Ref. \onlinecite{g4}, the simple choices $\mu_n = \mu_0/\ln 2$, a constant and
$w_n = w_{n-1}/\lambda$ give power-law relaxation. Here, this
situation is examined more fully.

The sum in Eq.\ (6) is rewritten as an integral using the above choices for
$w$ and $\mu$:
\bq
\Phi(t) = w_0 \int_0^\infty dn \lambda^{-n} \exp[(-t/\tau_0)e^{-n\mu_0}],
\eq
where from normalization at $t=0$, $w_0 = \ln\lambda$. This integral is
evaluated exactly in terms of the incomplete gamma function
\bq
\gamma(\nu, u) = \int_0^u dx \, x^{\nu-1}e^{-x}.
\eq
The result is 
\bq
\Phi(t) = \nu\left(\frac{\tau_0}{t}\right)^\nu\gamma(\nu,
t/\tau_0),
\eq
where $\nu \equiv \ln\lambda/\mu_0$.
For large values of its second argument, $\gamma(\nu,u)$ approaches
the complete gamma function $\Gamma(\nu)$. Thus, at large times
$\Phi(t)\propto t^{-\nu}$, as observed \cite{icko}. For small $u$,
$\gamma(\nu,u) \approx u^\nu/\nu - u^{\nu+1}/(\nu+1) +\ldots$, so
that for short times, $\Phi(t)$ crosses over to a slower $1-{\rm
const}(t/\tau_0)$ dependence. 
Temperature dependence is introduced in the model through the
temperature dependence of the fundamental relaxation time
$\tau_0(T)$; its inverse corresponds to the rate constant
$k_r(T)$ introduced in Ref.\ \onlinecite{icko}. Thus the behavior of Eq.\ (9) is
similar to the form
\begin{equation}
\Phi(t) = [1+k_r(T)\, t\,]^{-\nu}
\end{equation}
which was used in Fig.\ 3 of Ref.\ \onlinecite{icko} to fit the data.

The appearance of the experimental points at different
temperatures, in particular that the data are parallel (the same
power law for all $T$) at long times, suggests that a scaling
function could describe the experimental results. A general form is
\begin{equation}
\Phi(t, T) \propto T^{\alpha} G[t/\tau_0(T)],
\end{equation} Therefore, by rescaling the time by a parameter
$\tau_0(T)$ for each temperature the data for $\Phi(t,T)/T^{\alpha}$
would all fall on a single curve. The fact that the  power
law of the long time behavior is independent of $T$ implies that
the exponent
$\alpha=0$. If this rescaling is carried out, the temperature
dependence of the characteristic rescaling time $\tau_0(T)$ may be
determined. This rescaling for the data of Iben et al \cite{icko} is
shown in Fig.\ 2a.
\begin{figure}
\includegraphics[width=9cm]{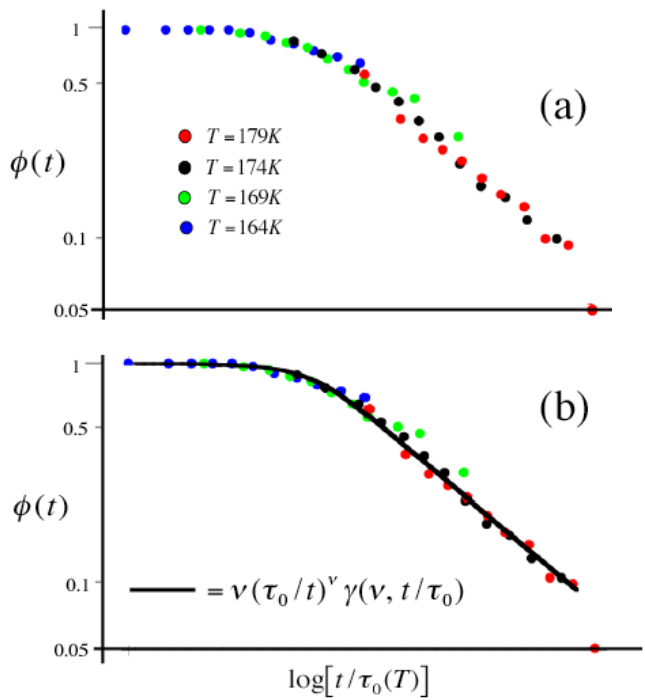}
\vskip -.1in
\caption{
(a) Rescaled data of Iben {\it et al}. \cite{icko}{\hskip .1in}
(b) Comparison of theory and rescaled data
}
\label{fig2}
\end{figure}


If in addition one has a theoretical expression which has a
scaling form, such as Eq. (9), then by collapsing the data onto the
theoretical curve, one obtains the numerical values of
$\tau_0(T)$. This is carried out for the hierarchical model of Eq.\
(9) in Fig.\ 2b. The value of $\nu$ in Eq.\ (9) is adjusted to
agree with the common large $t$ slope as seen in Fig.\ 2a. The
black curve in Fig.\ 2b is Eq.\ (9) with $\nu$ set equal to 0.28. The
values of $\tau_0$ for each of the four temperatures are plotted in Fig.\
3, where the solid curve is a fit using the
expression 
\begin{equation}
1/\tau_0(T) = k_0 \exp[-(T_0/T)^2]
\end{equation}
The result is $k_0 =
1.5\times 10^{16}\;{5\rm s}^{-1},\; T_0 = 1105$ K.\cite{cf} 
\begin{figure}[h]
\hskip -.7in\includegraphics[width=9cm]{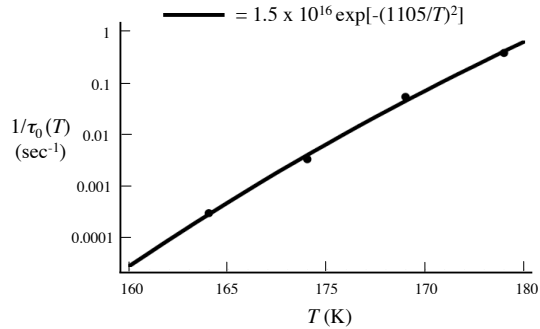}
\vskip -.3in
\caption{Fundamental relaxation rate}
\label{fig3}
\end{figure}
With only four points, a number of different functional forms might seem
equally good. In particular, an Arrhenius form works almost as well, but has
an unacceptably large pre-exponential factor. \cite{pnas} The form chosen
has been suggested by several authors
\cite{icko,bass}; it can describe diffusion in a random
potential of a form which mimics the potential surface of a protein.

As discussed in Ref.\ \onlinecite{icko}, the observed relaxation process most
likely involves substates having different conformational molecular
structures as well as different angles of the CO ligand with
respect to the heme normal. With this in mind, contact may be made
between the parameters of the above model and the actual
experiment. Recall that $\mu_0$ measures the number of ``spins" in
a given level which must arrive at a certain configuration before a
typical spin in the next level becomes unfrozen and can relax. The
physical picture of constrained movement of atoms which underlies
the model and its application to MbCO then implies that this number
should be around 3 to 5 or so. The number $\lambda$ measures the
geometric reduction of the fraction of spins which belong to level
$n$ as $n$ increases through slower and slower levels. $\Phi(t)$
has only a logarithmic dependence on $\lambda$ so it is reasonable
to take $\ln\lambda$ to be of order unity. This gives $\nu =
\ln\lambda/\mu_0\lesssim 1$. The fit value is $\nu = 0.28$,
i.e. slightly less than four spins, on average, combine to
unfreeze the degrees of freedom at the next level. This seems quite
reasonable.

Eq.\ (10), taken from Eq.\ (1) of Ref.\ \onlinecite{icko} is consistent with the
present Eq.\ (9) in that they agree in the limits of small and
large $t/\tau_0$. Therefore the fit by Iben {\it et al}. \cite{icko} using
Eq.\ (10) is also satisfactory. However, there is no physical
motivation for that form whereas in the present work the observed
non-exponential relaxation in MbCO is derived from a microscopic
model which incorporates strong-correlation constraints on the
relaxation of the molecule in a definite way.

What conclusions can be drawn from the present analysis? The scenario
that the model describes is one in which the primary relaxation event
represents the rate $1/\tau_0(T)$ at which a typical enthalpy barrier is
overcome. Since no further $T$-dependence is introduced, the conclusion is
that all subsequent relaxation events are ``slaved"\cite{pnas} to the primary
one and that they represent entropic conformational changes of the molecule.

A Los Alamos group has independently been analyzing the relaxation
processes in MbCO. \cite{pnas,lanl} Remarkably, they have reached
conclusions which are consistent with the present hierarchical model for
the relaxation of the CO vibration frequency. Namely, they argue that the
temperature dependence of the relaxation is governed by an activation
enthalpy for which the solvent is responsible. This determines the rate of the fastest
relaxation process - $1/\tau_0(T)$ in the hierarchical model. Subsequent
relaxations involve degrees of freedom of the protein and the hydration
shell and are governed by entropy barriers; they have the same temperature
dependence as the solvent fluctuation rate and are slower. This description
is precisely the same as that of the hierarchical model presented here.
The physically motivated scenario of the  successful hierarchical approach lends
support to the identification of the physical relaxation processes described in
Ref.\ 12.

The analysis presented here can be generalized to more complicated situations. For
example, the hierarchical rules could be modified to include simultaneous parallel
relaxation (``unslaved") processes as in the ligand rebinding referred to earlier, internal enthalpic barriers, reverse
constraints, and intra-level correlations. While other forms than Eq.\ (9) might
fit the MbCO data, within the hierarchical scenario the fact that the long time
behavior is a scalable power law practically forces the simple rules which were
used to obtain Eq.\ (9). The success of this approach suggests that a similar
picture and analysis can be relevant for other dynamical processes in biological
molecules. If so, insight can be obtained about the physical processes which
determine the relaxation phenomena. 


The author wishes to acknowledge helpful and often critical discussions with R.
Austin, S. Doniach, P. Fenimore, H. Frauenfelder, B. McMahon, B. Shklovskii, D. Stein. A
portion of this work was carried out during the author's participation in activities
of the Institute for Complex Adaptive Matter (ICAM). The hospitality of the Aspen
Center for Physics, where the research was begun and completed, is gratefully
acknowledged.




\begin{thebibliography}{99}
\bibitem{gold} V.I. Goldanskii, Yu.F. Krupyanskii and V.N. Flerov, Dokl. Akad. Nauk SSSR {\bf 272}, 978 (1983).
\bibitem{dls}  D.L. Stein, Proc. Natl. Acad. Sci. USA {\bf 82}, 3670 (1985).
\bibitem{icko} I.E.T. Iben, {\it {\it et al}.}.,
Phys. Rev. Lett. {\bf 62}, 1916 (1989).
\bibitem{shibata} Y. Shibata, A. Kurita, and T. Kushida, Biochemistry {\bf
38}, 1789-1801 (1999).
\bibitem{vf} H. Vogel, Phys. Z. {\bf 22}, 645 (1921); G.S.
Fulcher, J. Am. Ceram. Soc. {\bf 8}, 339 (1925).
\bibitem{g4} R.G. Palmer, D.L. Stein, E. Abrahams, and P.W.
Anderson, Phys. Rev. Lett. {\bf 53}, 958 (1984).
\bibitem{rha} R.H. Austin, K.W. Beeson, L. Eisenstein, and H. Frauenfelder, Biochemistry, {\bf 14}, 5355 (1975); Noam Agmon and J.J. Hopfield, J. Chem. Phys. {\bf 79}, 2042 (1983).
\bibitem{finite} One may ask whether taking an infinite sum in
Eq.\ (6) would be appropriate for a finite system. It can be
checked that for the analysis which follows for MbCO, taking a finite
sum over only 3 levels, say, gives less than a 10\% correction to the
relaxation function at long times.
\bibitem{bass} H. B\" assler, Phys. Rev.
Lett. {\bf 58}, 767 (1987); M. Gr\" unewald, {\it {\it et al}.}., Philos.
Mag. B {\bf 49}, 341 (1984); R. Zwanzig, Proc Natl. Acad. Sci. USA. {\bf
85}, 2029 (1988)>
\bibitem{cf} These numbers are of course rather close to the ones
determined in Ref.\ 1: $\nu = 0.26,\; k_0 = 10^{17},\; T_0 = 1130$ K
\bibitem{pnas} P.W. Fenimore, H. Frauenfelder, B.H. McMahon, and F.G. Parak, 
Proc. Natl. Acad. Sci. USA {\bf 99}, 16047 (2002).
\bibitem{lanl} P.W. Fenimore, H. Frauenfelder, B.H. McMahon, and R.D. Young,
Proc. Natl. Acad. Sci. USA {\bf 101}, 14408 (2004).
\end{thebibliography}
\end{document}